\documentclass[prd,a4paper,showpacs,nofootinbib]{revtex4}

\usepackage{epsfig}

\newcommand{\be}{\begin{equation}}
\newcommand{\ee}{\end{equation}}
\newcommand{\ba}{\begin{eqnarray}}
\newcommand{\ea}{\end{eqnarray}}
\newcommand{\nn}{\nonumber}

\newcommand{\simorderr}{\raisebox{-4pt}{$\, \stackrel{\textstyle <}{\sim} \,$}}
\newcommand{\lf}{\left}
\newcommand{\rg}{\right}

\begin{document}

\title{Compatibility of phenomenological dipole cross sections with the Balitsky-Kovchegov equation}

\author{Dani\"el Boer}
\email{D.Boer@few.vu.nl}
\affiliation{Department of Physics and Astronomy,
Vrije Universiteit Amsterdam, \\
De Boelelaan 1081, 1081 HV Amsterdam, The Netherlands}

\author{Andre Utermann}
\email{A.Utermann@few.vu.nl}
\affiliation{Department of Physics and Astronomy,
Vrije Universiteit Amsterdam, \\
De Boelelaan 1081, 1081 HV Amsterdam, The Netherlands}

\author{Erik Wessels}
\email{E.Wessels@few.vu.nl}
\affiliation{Department of Physics and Astronomy,
Vrije Universiteit Amsterdam, \\
De Boelelaan 1081, 1081 HV Amsterdam, The Netherlands}

\begin{abstract}

  Phenomenological models of the dipole cross section that enters in
  the description of for instance deep inelastic scattering at very
  high energies have had considerable success in describing the
  available small-$x$ data in both the saturation region and the
  so-called extended geometric scaling (EGS) region. We investigate to
  what extent such models are compatible with the numerical solutions
  of the Balitsky-Kovchegov (BK) equation which is expected to
  describe the nonlinear evolution in $x$ of the dipole cross section
  in these momentum regions. We find that in the EGS region the BK
  equation yields results that are qualitatively different from those
  of phenomenological studies. In particular, geometric scaling around
  the saturation scale is only obtained at asymptotic rapidities.  We
  find that in this limit, the anomalous dimension $\gamma(r,x)$ of
  phenomenological models approaches a limiting function that is
  universal for a large range of initial conditions. At the saturation
  scale, this function equals approximately $0.44$, in contrast to the
  value $0.628$ commonly used in the models. We further investigate
  the dependence of these results on the starting distribution, the
  small-$r$ limit of the anomalous dimension for fixed rapidities and
  the $x$-dependence of the saturation scale.
\end{abstract}


\maketitle


\section{Introduction}

At very high energy, deep inelastic scattering (DIS) of a virtual
photon off a proton can be described as the scattering of a color
dipole off small-$x$ partons, predominantly gluons, in the proton
\cite{Mueller:1989st}. The evolution in $x$ of the corresponding
dipole cross section has been the subject of many investigations.  The
earliest and best-known equation describing $x$-evolution at high
energy, i.e.\ small values of $x$, is the BFKL equation
\cite{Kuraev:1977fs,Balitsky:1978ic}. It is a linear evolution
equation based on resummation of logarithms in $1/x$, which are
dominant at small $x$. The BFKL equation predicts an exponential
growth of the dipole cross section as $x$ decreases, which potentially
violates unitarity.  Attempts to solve this problem have resulted in
the formulation of a variety of nonlinear evolution equations, the
earliest example of which is the GLR equation
\cite{Gribov:1984tu,Laenen:1995fh} for the gluon distribution. More
recently, a nonlinear evolution equation for the dipole cross section
has been obtained, the Balitsky-Kovchegov (BK) equation
\cite{Balitsky:1995ub,Kovchegov:1999yj}.  As a consequence of the
nonlinearity, the dipole cross section saturates with decreasing $x$,
thereby offering a resolution to the unitarity problem arising from
the BFKL equation.

The momentum scale at which nonlinearity becomes important and
saturation sets in is called the saturation scale $Q_s$. This scale 
grows as $x$ decreases, which means that if $Q_s$ becomes larger than 
1 GeV say, the coupling constant at $Q_s$ will be small enough for
small coupling methods to become applicable. 
In spite of this, even for the simplest nonlinear 
evolution equation --the BK equation-- analytical solutions have been 
found in specific cases only
\cite{Kovchegov:1999ua,Levin:1999mw,Munier:2003vc,Munier:2003sj,Marquet:2005qu,Marquet:2005ic,Kozlov:2005mm}.
However, numerical solutions to the BK equation have been
obtained in all momentum regions
\cite{Braun:2000wr,Braun:2001kh,Lublinsky:2001bc,Lublinsky:2001yi,Armesto:2001fa,Levin:2001et,Golec-Biernat:2001if,Enberg:2005cb}.
For a review of the properties of the numerical solutions cf.\
\cite{Stasto:2004rm}.

In the absence of exact solutions of the nonlinear evolution
equations, the dipole cross section has been modeled using both
theoretical considerations and experimental data. In the saturation region,
such models have been derived theoretically on the basis of
multiple rescattering of the dipole off the target, to various levels
of sophistication. In what is now commonly referred to as the Color
Glass Condensate (CGC) formalism \cite{MV,Iancu}, the small-$x$
gluons are described as a color field generated by the large-$x$
partons, which act as static sources. Scattering amplitudes are
expressed as two-point functions of lightlike Wilson lines that
are averaged over the possible source configurations, which
evolve with $x$. The averaged two-point functions satisfy a tower of
nonlinear evolution equations, the Balitsky-JIMWLK equations \cite{jimwlk},
coupling them to ever higher $n$-point functions. However, in a
mean field approximation the Balitsky-JIMWLK equations reduce to a
single evolution equation, the BK equation. The saturated state of
high-gluon density at small $x$ described by the CGC has been
under very active theoretical and experimental investigation in
recent years. Valuable further tests will be provided by future
colliders that are able to probe well into the saturation region,
such as the LHC and a possible electron-ion collider.

The phenomenological models for the dipole cross section that are used
to investigate the available HERA and RHIC data are constructed on the
basis of general theoretical arguments and contain only very few free
parameters. In this way satisfactory descriptions of several types of
processes have been obtained
\cite{GBW,Bartels:2002cj,IIM,KKT,DHJ1,DHJ2}.  The data thus seem to
confirm some general expectations concerning the dipole cross section,
in and above the saturation region. Here we want to investigate
whether the phenomenological models are compatible with the BK
equation. We will compare the numerical solutions of the BK equation
\cite{Enberg:2005cb} with a recent model for the dipole cross section
\cite{DHJ2} that is intended to describe the momentum and
$x$-dependence in a particular region above the saturation scale --the
extended geometric scaling region. Our conclusions will also apply to
the dipole models of \cite{KKT,IIM} in that region.  Regarding the
$x$-dependence it should be mentioned that the leading order BK
equation is known \cite{Triantafyllopoulos:2002nz,MuellerTr} to lead
to a faster evolution in $x$ ($Q_s^2(x) \sim 1/x^\lambda$ where
$\lambda \simeq 0.9$\footnote{The exact value of $\lambda$ depends on
$\bar{\alpha}_s=N_c/\pi\,\alpha_s$, $\lambda \simeq 0.9$ corresponding
to $\bar{\alpha}_s \simeq 0.2$.}) than the experimental data seem to
favor ($\lambda \simeq 0.3$). This discrepancy can be reduced by
introducing a running coupling constant and indeed in that case the
DIS data seems to be describable taking both LO BK evolution and, for
larger $Q^2$, DGLAP evolution into account
\cite{Gotsman:2002yy}. However, the consistency of including running
coupling in an equation that is of leading order in $\alpha_s$ is
questionable.  This indicates that going beyond the leading order BK
equation may be required in order to arrive at a proper quantitative
comparison. Here however, we will be mainly interested in the question
of whether the qualitative features of the models are reflected by the
solutions to the leading order BK equation, especially around the
saturation scale and at subasymptotic rapidities. 

The outline of this paper is as follows. In Section II we discuss the
properties of the phenomenological models of the dipole cross section
in detail. In Section III the BK equation is briefly reviewed. Section
IV is devoted to the study of the numerical solutions of the BK
equation, by means of a comparison with a general Ansatz for the
dipole cross section in terms of an anomalous dimension, in
particular, the one proposed in Ref.~\cite{DHJ2}. We also study the
$x$-dependence of the saturation scale with and without running
coupling and consider the dependence of the results on the initial
conditions. In Section V we summarize the main conclusions that can be
drawn from the comparisons we have performed.

\section{Phenomenological studies of the dipole cross section}
\label{pheno_sec} A very successful phenomenological study of experimental
data using a model for the dipole cross sections was performed by
Golec-Biernat and W\"{u}sthoff (GBW)~\cite{GBW}. The HERA data on
the DIS structure function $F_2$ at low $x$ ($x \simorderr 0.01$)
they analyzed\footnote{A further test of the dipole picture can be obtained 
from ratios of structure functions for which a bound has been derived recently 
\cite{Ewerz:2006an}.} could be described well by a dipole cross section
of the form $\sigma = \sigma_0 N_{GBW}(r_t,x)$, where the scattering
amplitude $N_{GBW}$ is given by 
\be
N_{GBW}({r}_t,x) = 1-\exp\left[-\frac{1}{4} r_t^2
Q_s^2(x) \right], \label{NGBW}
\ee
${r}_t$ denotes the transverse size of the dipole, 
$\sigma_0 \simeq 23\;{\rm mb}$
and the $x$-dependence of the
saturation scale is given by
\be
Q_s(x) = 1\,\mathrm{GeV}\,
\left(\frac{x_0}{x}\right)^{\lambda/2}, \label{Qsx2}
\ee
with $x_0 \simeq 3 \times 10^{-4}$ and $\lambda \simeq 0.3$. 
Although here
we discuss DIS off the proton ($A=1$), we mention that for nuclear
targets $Q_s^2$ increases as $A^{1/3}$.

The above scattering amplitude displays a characteristic feature
called geometric scaling. This means that it depends on $x$ and $r_t^2$
($r_t$ being the Fourier conjugate of $k_t=Q$) through the
combination $r_t^2Q_s^2(x)$ only. A model independent analysis of
the DIS data from HERA shows that the low-$x$ data display
geometric scaling for all $Q^2$~\cite{Stasto:2000er}, even though
the GBW model, (\ref{NGBW})-(\ref{Qsx2}), was found to be
inconsistent with newer, more accurate data~\cite{Bartels:2002cj}
at large $Q^2$. In Ref.\ \cite{Bartels:2002cj} a modification of
the GBW model was proposed which includes DGLAP evolution, as
required to fit the $Q^2 > 20$ GeV$^2$ data. Alternatively, in
Ref.\ \cite{Gotsman:2002yy} the dipole scattering amplitude from the
GBW model was replaced by a numerical solution to the leading order 
BK equation with running coupling and 
with the addition of a correction that satisfies DGLAP evolution
in order to also describe the short distance behavior of $N$
correctly. As emphasized in Ref.~\cite{IIM} (IIM), the solution to
the BK equation (with running coupling) 
by itself only provides a satisfactory fit for
relatively low $Q^2$, up to a few GeV$^2$.

As mentioned, the DIS data show geometric scaling for $x<0.01$ and all
$Q^2$ (this was recently confirmed in Ref.\ \cite{Gelis:2006bs}), but
the data above and below $Q^2/Q_s^2(x) \approx 1$ behave differently
as a function of $Q^2/Q_s^2(x)$ \cite{Stasto:2000er}.  On the
basis of theoretical considerations one expects different geometric
scaling behavior at low and high $Q^2/Q_s^2$ (a Gaussian-like function
of $Q^2/Q_s^2$ at low $Q^2/Q_s^2$ and a power law fall-off at high
$Q^2/Q_s^2$). Moreover, one expects an intermediate region, commonly
referred to as the extended geometric scaling (EGS)
region~\cite{LevinTuchin,IIM2,MuellerTr}, starting from $Q_s$,
extending to a scale $Q_{gs}$, where the geometric scaling
behavior of the saturation region holds approximately. 
This EGS region grows with decreasing $x$. 

An estimate of $Q_{gs}$ has been obtained from the solution of the
LO-BFKL evolution equation expanded to second order around the
saturation saddle point~\cite{IIM2}. Here $Q_{gs}$ is taken to be the scale 
at which the violation of the geometric scaling behavior that holds at
$Q_s$ becomes of the same order as the leading order scaling contribution. 
One then finds $Q_{gs}(x) \sim Q_s^2(x)/\Lambda$, where $\Lambda$ is a nonperturbative
scale of order $\Lambda_{\rm QCD}$. A similar result for $Q_{gs}$ can
be obtained from estimating the transition point between the LLA and
DLA saddle points~\cite{JK}. We emphasize that at
$Q_{gs}$ one does not necessarily observe large deviations
from geometric scaling, one may also reach a region with a
different geometric scaling behavior.

The theoretical studies of the EGS region have led IIM~\cite{IIM} to propose a
modification to the GBW model so as to obtain a satisfactory fit
to DIS data in both the saturation region and the EGS region,
without including DGLAP evolution at large $Q^2$. In the
same spirit, KKT \cite{KKT} have considered a modification of the
GBW model in the EGS region in order to obtain a good description
of the $p_t$ spectra of hadron production at RHIC in $d$-$Au$
collisions (for a review of saturation physics and $d$-$Au$ collisions
cf.\ \cite{JK}). Here we will restrict the discussion to the later
modifications of the KKT model made by DHJ \cite{DHJ2}, but our
conclusions for the EGS region will apply to the IIM and KKT models
as well. We note that the saturation scale of the IIM model is to 
be multiplied by two in order to compare with the results obtained
below. Other dipole models are discussed in 
Refs.\ \cite{Goncalves:2006yt,deSantanaAmaral:2006fe}. 

The model
proposed by DHJ offers a good description of the $p_t$ spectra of
hadron production at RHIC in $d$-$Au$ collisions in both the
midrapidity and forward regions, and even, as has been shown
recently \cite{BDH}, the $p$-$p$ data in the very forward rapidity
region.

The modified dipole scattering amplitude in the EGS region is
given by~\cite{KKT,DHJ1,DHJ2}:
\be
N({r}_t,x) =
1-\exp\left[-\frac{1}{4}(r_t^2 Q_s^2(x))^{\gamma(r_t,x)}\right].
\label{Ngamma}
\ee 
The exponent $\gamma$ is usually referred to as the ``anomalous
dimension'', although the connection of $N$ with the gluon
distribution may not be clear for all cases considered
below.  This anomalous dimension $\gamma$ changes the evolution in
$x$, thus governing the possible violation of geometric scaling for
scales between $Q_s$ and $Q_{gs}$. The unitarized form of $N$ is
important only for scales around and below $Q_s$. For larger scales
one can use to good approximation 
\be 
N({r}_t,x) \approx
\frac{1}{4}(r_t^2 Q_s^2(x))^{\gamma(r_t,x)},
\label{NgammaApprox}
\ee
which allows one to make a connection with
the DGLAP region\footnote{More specifically, at
low $x$ and high $Q^2$ a logarithmic evolution in both these
variables may be required.}, 
where $\tilde{N}_{{\rm DGLAP}} (k_t,x) \propto
k_t^{-4}$ ($\tilde{N}$ being the Fourier transform defined in Eq.\
(\ref{Ntildek})), which is the tail behavior determined by one gluon 
($t$-channel) exchange. Hence, the ``DGLAP'' limit of $\gamma$ would be
$\gamma\to1$.

DHJ (following in part IIM and KKT) impose the following
requirements to determine a parameterization of $\gamma$ in the
EGS region:
\begin{enumerate}
\item For any fixed $x$, in the limit $r_t\to0$ $\gamma$ should approach
1. 
\item At the saturation scale, $\gamma$ should be
constant, so that $N$ displays exact geometric scaling:
$\gamma(r_t=1/Q_s,x)=\gamma_s$. The constant $\gamma_s$ is chosen
to be $\simeq 0.628$.
\item If
one writes $\gamma=\gamma_s+\Delta\gamma$, then $\Delta\gamma$
should decrease as $1/y$ for $y\to\infty$ at fixed $r_t^2 Q_s^2$.
\item At large, but fixed, rapidity $y$, the extent of the
geometric scaling window should be consistent with the estimated
scale $Q_{gs}=Q_s^2/\Lambda_{\rm QCD}$.
\end{enumerate}
A few comments on these requirements are in order. Requirement 1 is
imposed in order to reproduce the ``DGLAP'' limit $N \sim r_t^{2}$.
Requirement 2 and 3 were motivated by properties of the BFKL equation 
\cite{IIM}. The
idea is that since in the saddle point approximation the solution of the
BFKL equation exhibits the property of geometric
scaling, one can use it as an approximation to the solution of the
BK equation near the saturation scale $Q_s$, provided that one
imposes an appropriate boundary condition at $Q_s$
\cite{IIM2,MuellerTr,Triantafyllopoulos:2002nz}, commonly referred to as
a saturation boundary condition. Requirement 4
\cite{DHJ2} is of a more phenomenological nature, and determines
the speed with which $\gamma$ increases from $\gamma_s\simeq
0.628$ at $Q_s$ to a value near 1 at $Q_{gs}\approx
Q_s^2/\Lambda_{\rm QCD}$ (which is a function of $y$ through
$Q_s$). As can be seen from e.g. Ref.\ \cite{JK}, requirement
3 is a prerequisite for requirement 4.

Following the above requirements, the anomalous dimension
$\gamma(r_t,x)$ of DHJ is parameterized as \cite{DHJ2} 
\be
\gamma(r_t,x) =
\gamma_s + (1-\gamma_s)\, \frac{\log(1/r_t^2Q_s^2(x))}{\lambda
y+d\sqrt{y}+\log(1/r_t^2Q_s^2(x))}, \label{gammaparam}
\ee
where
$y=\log x_0/x$ is minus the rapidity of the target parton. The
saturation scale $Q_s(x)$ and the parameter $\lambda$ are taken
from the GBW model, as given in Eq.\ (\ref{Qsx2}). Requirement 4 is
satisfied, which can be seen as follows: for large, but fixed
rapidity ($y \gg (d/\lambda)^2$), the EGS region roughly extends to $\log
Q_{gs}^2/Q_s^2 \sim \lambda y$, which implies that numerically
$Q_{gs} \approx Q_s^2/\Lambda_{\rm QCD}$. The
parameterization~(\ref{gammaparam}) includes subleading
corrections $\sim d\sqrt{y}$ which govern geometric scaling
violations at subasymptotic rapidities. For $d\simeq1.2$ a good
description of $d$-$Au$ data from RHIC over a considerable 
rapidity range, from mid- to forward rapidities, 
was obtained~\cite{DHJ2}. This parameter will play no role in our
discussion, as we do not aim to obtain the best possible
expression for either $N$ or $\gamma$, but rather we want to
investigate whether the above type of parameterization is 
compatible with the BK equation.

Two further aspects of DHJ's Ansatz~(\ref{Ngamma}) for $N$ with
the anomalous dimension of Eq.\ (\ref{gammaparam}) should be
mentioned. First, in order to calculate observables (and to connect to
the BK equation), one needs to formulate the dipole scattering amplitude
in momentum space by performing a Fourier transform. To
simplify this procedure, DHJ (following a similar step by KKT)
replaced $\gamma(r_t,x)$ by $\gamma(1/k_t,x)$. This replacement should be
viewed as an approximation that becomes better as $k$ increases 
or equivalently, as $r$ decreases. This simplifying approximation
could be regarded as part of the Ansatz. Second, quarks and
gluons are described by different scattering amplitudes, since
they couple differently to the gluons. The above
Ansatz~(\ref{Ngamma}) was applied by DHJ to gluons and is referred to as
$N_A$. The corresponding expression for quarks is called $N_F$ and
is obtained from $N_A$ by the replacement $Q_s^2 \to (C_F/C_A)
Q_s^2=(4/9) Q_s^2$. In the comparison to the GBW model, which is a model
for $N_F$, one should take into account this factor, leading
to $Q_s(x_0) = 1$ GeV instead of 0.67 GeV.

Eq.\ (\ref{gammaparam}) was intended to parameterize $\gamma$ in
the geometric scaling window only, so the question arises what forms
$N$ takes in the saturation region, where $r_t > 1/Q_s$. 
For instance, it has been suggested \cite{BDH}
to be of the form (\ref{Ngamma}) with $\gamma$ a constant, equal to 
$\gamma_s$, so that the Ansatz for $N$
displays exact geometric scaling throughout the saturation region. 
However, this idea has the drawback that
the resulting $N$ is not a smooth function of $r_t$, a problem also
present in the IIM model. Its Fourier
transform would therefore show artificial oscillations. To resolve
this issue, it seems better to use the BK equation (or other, more
appropriate, evolution equations) to determine the behavior of $N$
in the saturation region. The goal of this paper is to study in
more detail the relation between the DHJ Ansatz and the BK
equation. We use the ``BKsolver'' program \cite{BKSolver} to obtain a
numerical solution to the BK equation and study whether the  
dipole profile of DHJ and similar models is consistent with this solution. 
Also, we will address the behavior of $N$ in the
saturation region as dictated by the BK equation.
More specifically, we will investigate whether the requirements 2
and 3 on $\gamma$, which were motivated by the BFKL equation with 
a saturation boundary condition, are
consistent with the nonlinear evolution of the BK equation.
\section{Balitsky-Kovchegov equation}
\label{sect_BK}
The BK equation for $N$ (more specifically, for $N_F$) reads 
\cite{Balitsky:1995ub,Kovchegov:1999yj}
\ba
\frac{\partial N\lf((\vec{x}_t-\vec{y}_t)^2,x\rg)}{\partial y} & = &
\frac{\bar{\alpha}_s}{2\pi} \int d^2 z_t
\frac{(\vec{x}_t-\vec{y}_t)^2}{(\vec{x}_t-\vec{z}_t)^2
(\vec{y}_t-\vec{z}_t)^2} \Big[
N\lf((\vec{x}_t-\vec{z}_t)^2,x\rg)+N\lf((\vec{z}_t-\vec{y}_t)^2,x\rg) \nn \\[2 mm]
&& \mbox{} \hspace{3.5 cm}
-N\lf((\vec{x}_t-\vec{y}_t)^2,x\rg)-N\lf((\vec{x}_t-\vec{z}_t)^2,x\rg)
N\lf((\vec{z}_t-\vec{y}_t)^2,x\rg) \Big].
\ea
Here $\bar{\alpha}_s = \alpha_s N_c/\pi$. We will not consider the impact parameter dependence of $N$ in this paper.

Two types of Fourier transform are often considered in this context:
\ba
\tilde{N}(k,x) & \equiv & \int \frac{d^2 r_t}{2\pi} e^{i \vec{k}_t
    \cdot \vec{r}_t} N(r,x),
\label{Ntildek} \\
{\cal N}(k,x) & \equiv & \int \frac{d^2 r_t}{2\pi} e^{i \vec{k}_t
    \cdot \vec{r}_t} \frac{N(r,x)}{r^2},
\label{Nk} 
\ea 
where $r^2=\vec{r}_t^2$ and $k^2=\vec{k}_t^2$. The
first (regular) Fourier transform is the one that enters the cross
section descriptions, such as the one considered by DHJ
\cite{DHJ1,DHJ2}. The second transform, which includes the
additional factor $1/r^2$, is more directly related to the
(unintegrated) gluon distribution, as explained in detail in Ref.\
\cite{Kharzeev:2003wz}. In terms of ${\cal N}$, the BK equation can be
very compactly written as \be
\partial_Y {\cal N} = \chi(-\partial_L) {\cal N} - {\cal N}^2,
\label{BKforcalN} \ee where $Y=y \bar{\alpha}_s$,
$L=\log(k^2/k_0^2)$, for some arbitrary scale $k_0$, and
$\chi(\gamma)$ is the BFKL kernel, \be \chi(\gamma) = 2
\psi(1)-\psi(\gamma)-\psi(1-\gamma).
\ee

In the limit of large $k$ ($\gg Q_s$), or equivalently, small $r$, the
nonlinear BK equation (\ref{BKforcalN}) reduces to the linear BFKL
equation.  In the saddle point approximation, the solution of the BFKL
equation behaves as $N(r,x) \sim r^{2\gamma'}$ for small $r$, hence
the exponent $\gamma'$ determines the small-$r$ limit of the anomalous
dimension $\gamma$ through Eq.\ (\ref{NgammaApprox}). Therefore,
$\gamma$ is determined by either the BFKL saddle point $\gamma_{s.p.}$
or the initial condition: if in the limit $r \to 0$ the starting
distribution has the property $N(r,x_0) \sim r^{2\gamma_0}$, then for
any fixed $x$ the asymptotically small $r$ behavior either remains
$r^{2\gamma_0}$ if $\gamma_0 < \gamma_{s.p.}$, or becomes
$r^{2\gamma_{s.p.}}$ if $\gamma_0 \geq \gamma_{s.p.}$. We will
demonstrate this observation explicitly below.  For finite $x$, the
BFKL saddle point approaches $\gamma_{s.p.} \approx 1$ in the limit of
small $r$, cf.\ e.g.\ \cite{IIM2}.  Therefore, any solution to the BK
equation will only satisfy requirement 1 of DHJ if the starting
distribution in the small-$r$ limit has the property that $\gamma_0
\geq 1$.

Around the saturation scale, the solution of the BK equation has been
approximated by a solution of the BFKL equation which has been
required to satisfy a saturation boundary condition
\cite{IIM2,MuellerTr,Triantafyllopoulos:2002nz}. This leads to the
aforementioned saddle point at saturation, $\gamma_s \approx 0.628$,
which is the solution of the equation
$\chi(\gamma_s)/\gamma_s=\chi'(\gamma_s)$ \cite{Gribov:1984tu,IIM2}.
However, unlike in the small-$r$ limit where the BFKL solution and
Eq.\ (\ref{Ngamma}) reduce to the same form, around the saturation
scale there is no clear correspondence between this saddle point and
the anomalous dimension of Eq.\ (\ref{Ngamma}).

Alternatively, one can first approximate the BK equation
(\ref{BKforcalN}) by expanding the kernel $\chi$ around the saddle
point, and subsequently solve the resulting equation analytically.
This has been done in Refs.\
\cite{Munier:2003vc,Munier:2003sj,Marquet:2005qu,Marquet:2005ic} where
a traveling wave solution has been obtained that is valid in a finite
range of $k$, which grows with the rapidity. The value $\gamma_s
\approx 0.628$ enters the analytic expressions for ${\cal N}(k,x)$ of
the traveling wave solutions\footnote{For initial conditions with
$\gamma_0<\gamma_s$, $\gamma_s$ is replaced with $\gamma_0$ in the
traveling wave expression.}. However, also for this solution
$\gamma_s$ does not correspond directly to the anomalous dimension of
Eq.\ (\ref{Ngamma}) at the saturation scale, since there is no direct
relation between $N(r,x)$ and ${\cal N}(k,x)$ at the respective points
$r=1/Q_s$ and $k=Q_s$.

We conclude that requirement 2 of DHJ, which sets $\gamma$ at the
saturation scale equal to $\gamma_s \approx 0.628$, does not follow
from these analytical results.  For completeness we mention that other
analytical aspects of the solutions of the BK equation, such as the
small-$k$ limit, have been investigated as well
\cite{Kovchegov:1999ua,Levin:1999mw,Kozlov:2005mm}.

\begin{figure}[htb]
\centering
\includegraphics*[width=86mm]{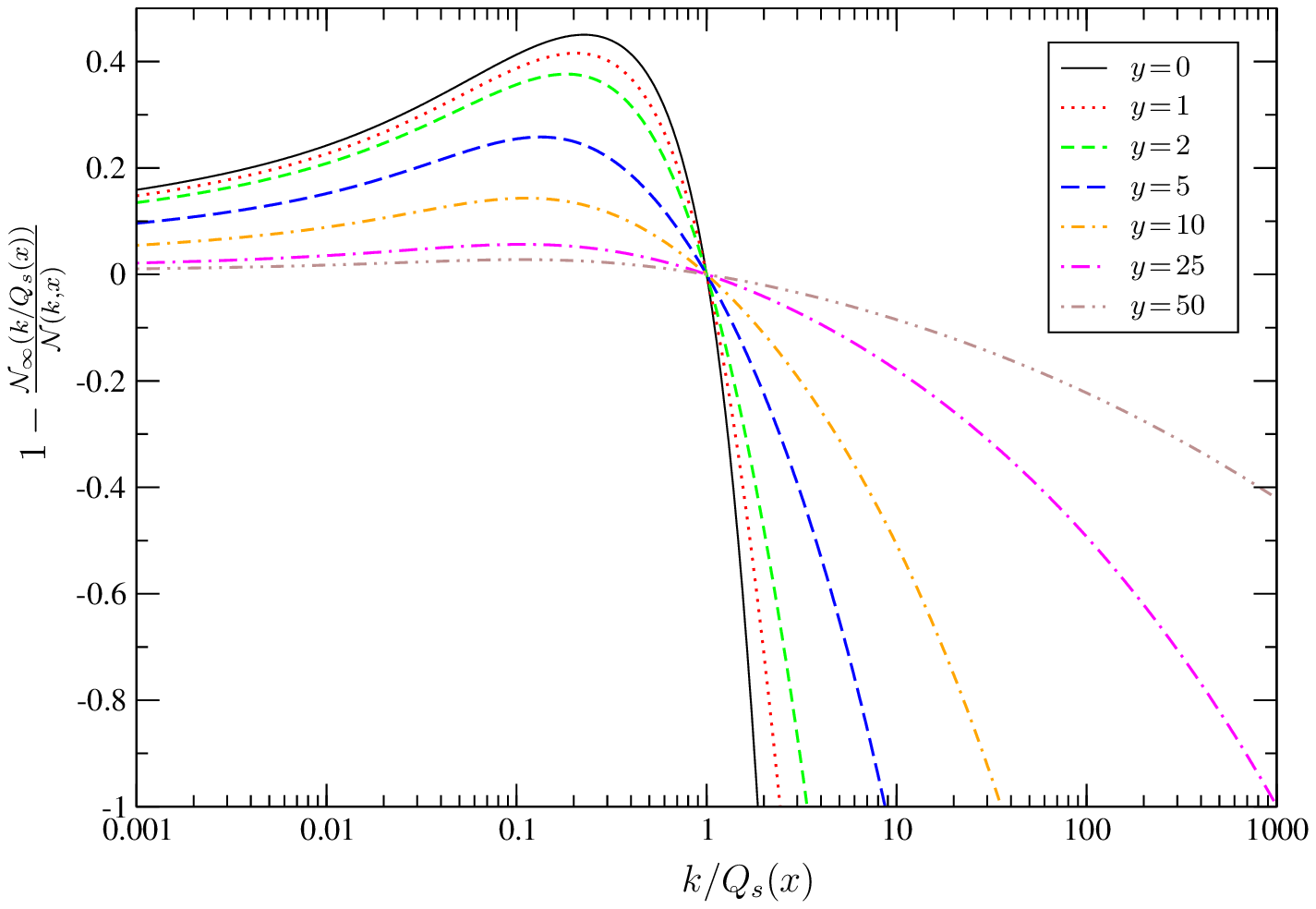}
\includegraphics*[width=86mm]{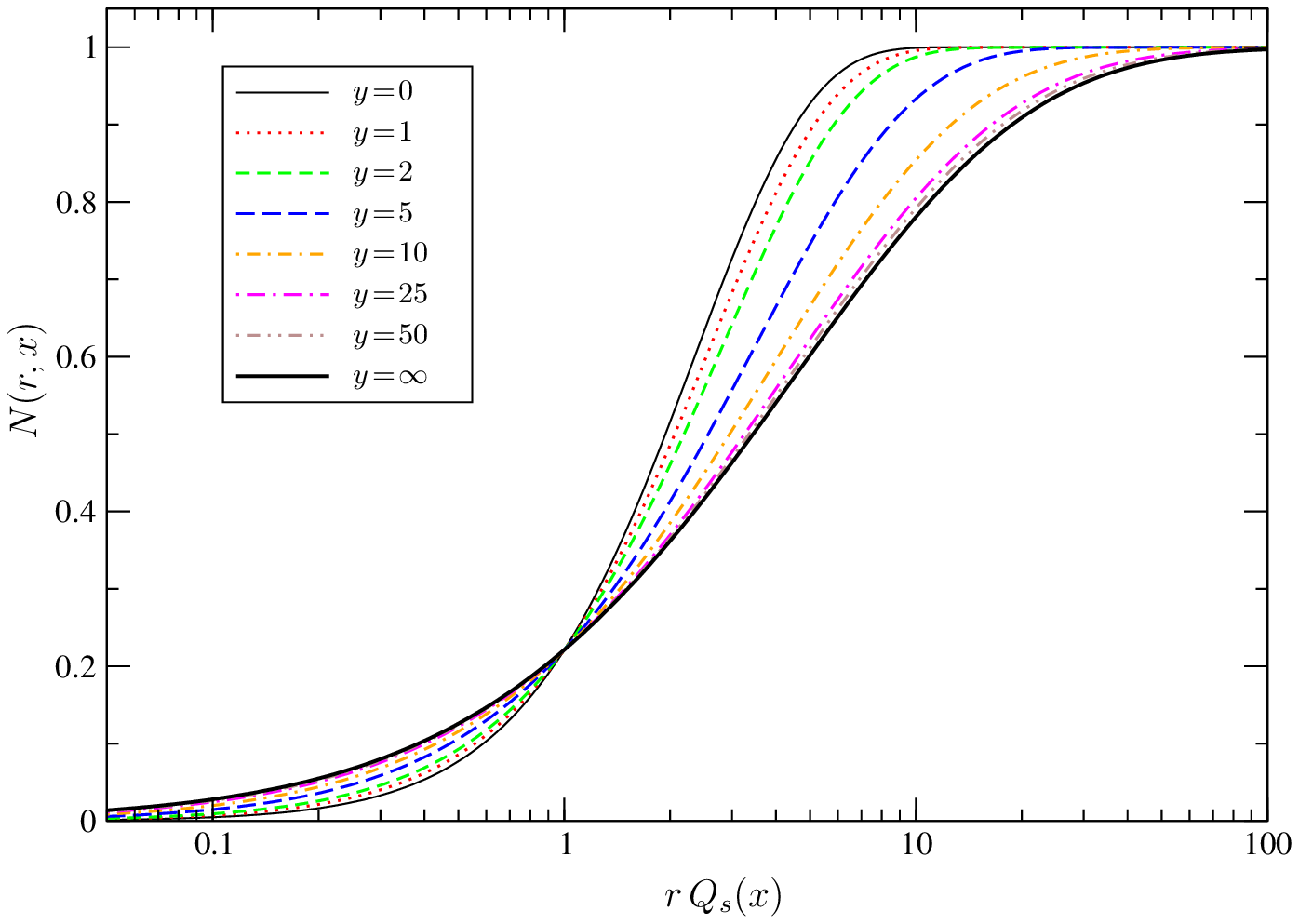}
\caption{\small\label{ncal_fig}a) The relative difference $1-{\cal N}_\infty(k/Q_s(x))/{\cal N}(k/Q_s(x),x)$ 
between the function ${\cal N}(k,x)$ (\ref{Nk}) following from the BK 
equation and its geometrically scaling asymptotic 
$\lim_{x\to  0}{\cal N}(k/Q_s(x),x)={\cal N}_\infty(k/Q_s(x))$ 
for various rapidities  $y=\log x_0/x$. \\
b) The dipole scattering amplitude $N(r,x)$ resulting from a Fourier 
transformation (\ref{Fourier}) of ${\cal N}(k,x)$ as a function of $r Q_s(x)$ 
for various rapidities  $y=\log x_0/x$.}
\end{figure}

Here, we will restrict ourselves to comparing the phenomenological DHJ
model with numerical solutions.  Such solutions have been obtained by
several groups
\cite{Braun:2000wr,Braun:2001kh,Lublinsky:2001bc,Lublinsky:2001yi,Armesto:2001fa,Levin:2001et,Golec-Biernat:2001if,Enberg:2005cb},
even including impact parameter dependence
\cite{Golec-Biernat:2003ym,Gotsman:2004ra,Albacete:2004gw,Marquet:2005zf}.
The code employed by the authors of Ref.\
\cite{Enberg:2005cb} to numerically solve the BK equation
has been made publicly available \cite{BKSolver}.
We will use this code in our subsequent investigation of
the DHJ Ansatz and its possible extension to the saturation region.
But first we want to point out that these numerical solutions do not
show geometric scaling for finite rapidities. The deviations can
become significant even in the saturation region at moderate $y$.
As is known, in
the limit of large rapidities, geometric scaling is
recovered; ${\cal N}(k,x)$ is given by a function ${\cal
  N}_\infty$ that depends on $k/Q_s(x)$ only. Fig.\ \ref{ncal_fig}a
shows the relative difference between ${\cal N}$ and the asymptotic
solution ${\cal N}_\infty$ as a function of $k/Q_s(x)$. The saturation scale
$Q_s(x)$ is defined as the value of $k$ where ${\cal N}$ equals some arbitrary
fixed value for which ${\cal N}^2$ becomes non-negligible. Based on our
study in Sec. IV.B we have chosen ${\cal N}(Q_s(x),x)=0.2$. Due to this
definition ${\cal N}-{\cal N}_\infty$ vanishes at the saturation
scale. Clearly, the scaling violations in the saturation
region are considerable for finite rapidities. Fig.\
\ref{ncal_fig}b shows the dipole scattering amplitude $N(r,x)$ in
coordinate space as it follows by a Fourier transformation
(\ref{Fourier}) from ${\cal N}(k,x)$. As expected, $N(r,x)$
vanishes for small $r$ and approaches 1 in the saturation region $r\gg
1/Q_s(x)$. Like in momentum space $N(r,x)$ approaches a limiting curve
that depends only on $r Q_s(x)$ and geometric scaling is recovered.
Throughout this paper we have chosen a constant coupling 
$\bar{\alpha}_s=N_c/\pi\,\alpha_s\approx 0.2$.

\section{Anomalous dimension from the Balitsky-Kovchegov equation}

\subsection{Anomalous dimension as a function of $r$ and $x$}
\begin{figure}[htb]
\centering
\includegraphics*[width=110mm]{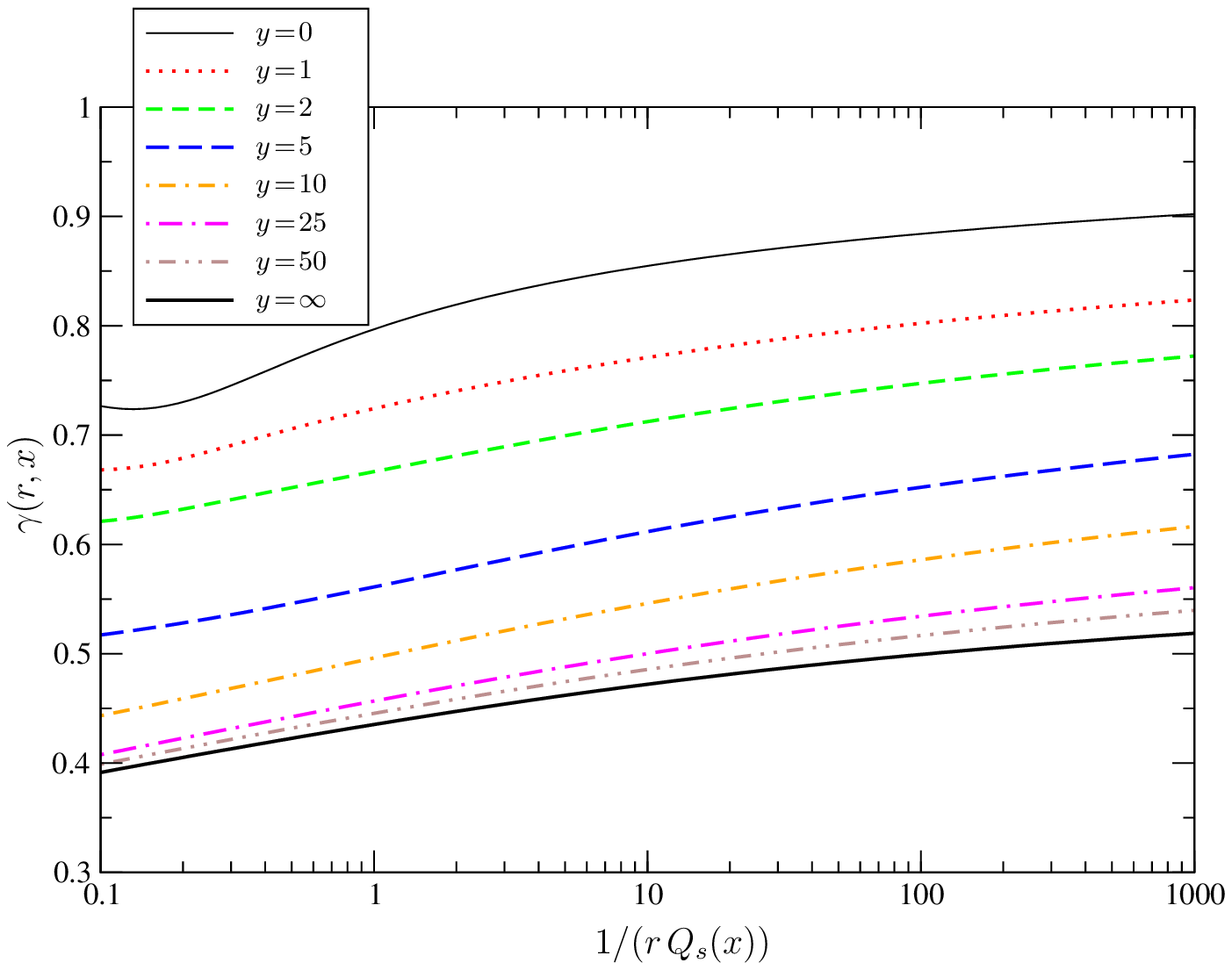}
\caption{\small\label{gammar1_fig} $\gamma(r,x)$ resulting from
  the relations (\ref{Fourier}) and  (\ref{gammar}) as
  a function of $1/(r Q_s(x))$ and $y=\log x_0/x$.}
\end{figure}

The BKsolver program \cite{BKSolver} provides numerical values for
the amplitude ${\cal N}(k,x)$ as a function in momentum space. In
order to use this solution of the BK equation to constrain
$\gamma(r,x)$, one first has to find $N(r,x)$ by Fourier
transforming back to coordinate space:
\begin{equation}
  N(r,x)=r^2\int\frac{d^2
    k_t}{2\pi}\:e^{-i\vec{k}_t\cdot\vec{r}_t}\,\mathcal{N}(k,x)
=r^2\int_0^\infty dk\,k\:{J}_0(k r)\,\mathcal{N}(k,x)\,.\label{Fourier}
\end{equation}
Using the Ansatz (\ref{Ngamma}) one can extract $\gamma(r,x)$ from
the resulting $N(r,x)$,
\begin{equation}
 \gamma(r,x)=\frac{\log\lf[\log\lf[\frac{1}{(1-N(r,x))^4}\rg]\rg]}{
\log[r^2\,Q_s^2(x)]}\,.
\label{gammar}
\end{equation}
This equation requires as a separate input the value of $Q_s(x)$,
which can be found as follows. One can fix $Q_s(x)$ in such a way
that at the saturation scale the amplitude obtained from the BK
equation (\ref{gammar}) equals the Ansatz (\ref{Ngamma}), which at
$Q_s$ becomes independent of $\gamma$,
\begin{equation}
 N(r=1/Q_s,x)=1-\exp\lf[-1/4\rg] \approx 0.22\,.
\label{fixsat}
\end{equation}
Equating this fixed value with the general relation (\ref{Fourier})
allows one to calculate the saturation scale $Q_s(x)$ by solving
the following equation,
\begin{equation}
 \frac{1}{Q_s^2(x)}\int_0^\infty dk\,k\:{J}_0(k/Q_s(x))\,\mathcal{N}(k,x)
=1-\exp\lf[-1/4\rg]\,.
\label{Qsdef1}
\end{equation}
Combining the resulting values of the saturation scale with Eq.\
(\ref{gammar}), we obtain a numerical result for $\gamma(r,x)$,
which is shown in Fig.~\ref{gammar1_fig} as a function of $1/(r
Q_s(x))$. This means that the saturation region is located to the
left of 1 on the horizontal axis, so that it is easier to compare
with later figures that display related quantities as a function
of $k/Q_s$. Note that although the Ansatz (\ref{Ngamma}) was
specifically intended to describe $N$ in the EGS region, we also
show $\gamma$ in the saturation region.

The resulting $\gamma(r,x)$ has the following features:
\begin{enumerate}
\item For $r \to 0$, $\gamma(r,x)$ asymptotically approaches 1.
\item At the saturation scale, $\gamma(r,x)$ is not a constant.
\item For decreasing $x$, $\gamma(r,x)$ approaches a limiting curve, 
indicated in Fig.~\ref{gammar1_fig} by $y=\infty$. 
Hence, after a longer evolution one indeed recovers geometric scaling.
\end{enumerate}

The fact that for small distances $\gamma$ 
asymptotically approaches 1 is understandable from the BK
equation, since in this limit it reduces to the BFKL equation.  
As mentioned before, in the limit of small distances, the solution to the BFKL 
equation is dominated by either the saddle point or the initial condition, 
both leading to $\gamma\to1$, since here we use the MV model (\ref{Ansatz2}) 
as the initial condition.

Requirement 2 of the DHJ Ansatz is clearly not satisfied by the
numerical solution of the BK equation at the saturation scale.
Instead of writing $\gamma(r,x)=\gamma_s + \Delta \gamma(r,x)$, it
seems more natural to consider the following split-up:
\begin{equation}
 \gamma(r,x)=\gamma_\infty(rQ_s(x))+\Delta\gamma(r,x)\,,
\end{equation}
where it turns out that, similar to requirement 3, $\Delta\gamma(r,x)$
decreases as $1/y$ for $y \to \infty$ and fixed $rQ_s(x)$. At the
saturation scale $\gamma$ is given in the small-$x$ limit by,
\begin{equation}
\lim_{x\to 0} \gamma(r=1/Q_s(x),x)=\gamma_\infty(1)\approx 0.44\,,
\label{gammars}
\end{equation}
a value that is significantly below $\gamma_s=0.628$. 
 
It would be interesting to see whether the numerically obtained
$\gamma$ could fit the available data as well, which does not seem unlikely
since KKT employed $\gamma_s =0.5$ and were able to fit $d$-$Au$ data, 
but this is not the aim of the present paper, where we instead focus on the 
comparison of existing models for the dipole cross section with the numerical
solution of the BK equation. Moreover, a word of caution has to be
added about the results for small values of $y$ (below $y=5$, say): the
numerical solutions are obtained after a choice of a starting
distribution at $y=0$ ($x=x_0$). For ``short'' evolutions (small $y$-values),
the properties of the starting distribution are still visible in the
result, i.e.\ not only in the small-$r$ limit. 
The starting distribution should be appropriate to the choice of $x_0$, 
which itself should be small enough for BK evolution to be applicable. 
If $x_0$ is 
sufficiently small, or alternatively if $A$ is sufficiently large, one can 
take for 
instance the McLerran-Venugopalan (MV) model \cite{MV}. One could also 
consider taking the DHJ model, which however is restricted to the EGS region, 
therefore requiring some extrapolation to the entire $r$ or $k$ range. For 
larger values of 
$x_0$ one may consider using one of the standard parton distributions, 
although these are not defined over the whole range either.

Due to this problem, 
throughout this paper we have used the MV model as a starting
distribution\footnote{There are many ways in which the infrared divergence
that occurs in the MV model can be regularized; the
replacement of $\log(1/(r^2\Lambda^2)) \to \log(e+1/(r^2\Lambda^2))$
is one option. In Ref.\ \cite{Jalilian-Marian:1996xn} it is shown
that the results do not depend much on the choice of regularization.}:
\begin{equation}
N(r,x)=1-\exp\lf[-\frac{1}{4}(r^2Q_s^2(x)) \log(e+1/(r^2\Lambda^2))\rg]\,,
\label{Ansatz2}
\end{equation}
which is one of the distributions considered in 
Ref.\ \cite{Enberg:2005cb}. In Sec IV.E we show some results for another
choice of starting distribution, to illustrate the influence of the initial condition. 
The dependence of the numerical results on the starting distribution has also 
been investigated in Ref.\ \cite{Enberg:2005cb}, to which we refer for additional information. 

Hence, for small rapidities ($y \simorderr 5$) one should avoid drawing too 
strong conclusions from a comparison of the DHJ parameterization with 
the numerical results that were obtained with the MV starting distribution. 
Nevertheless, we will be able to draw some qualitative conclusions also
for lower $y$-values from our results in Sec IV.B.

\subsection{The replacement $\gamma(r,x) \to \gamma(1/k,x)$}

\begin{figure}[htb]
\centering
\includegraphics*[width=110mm]{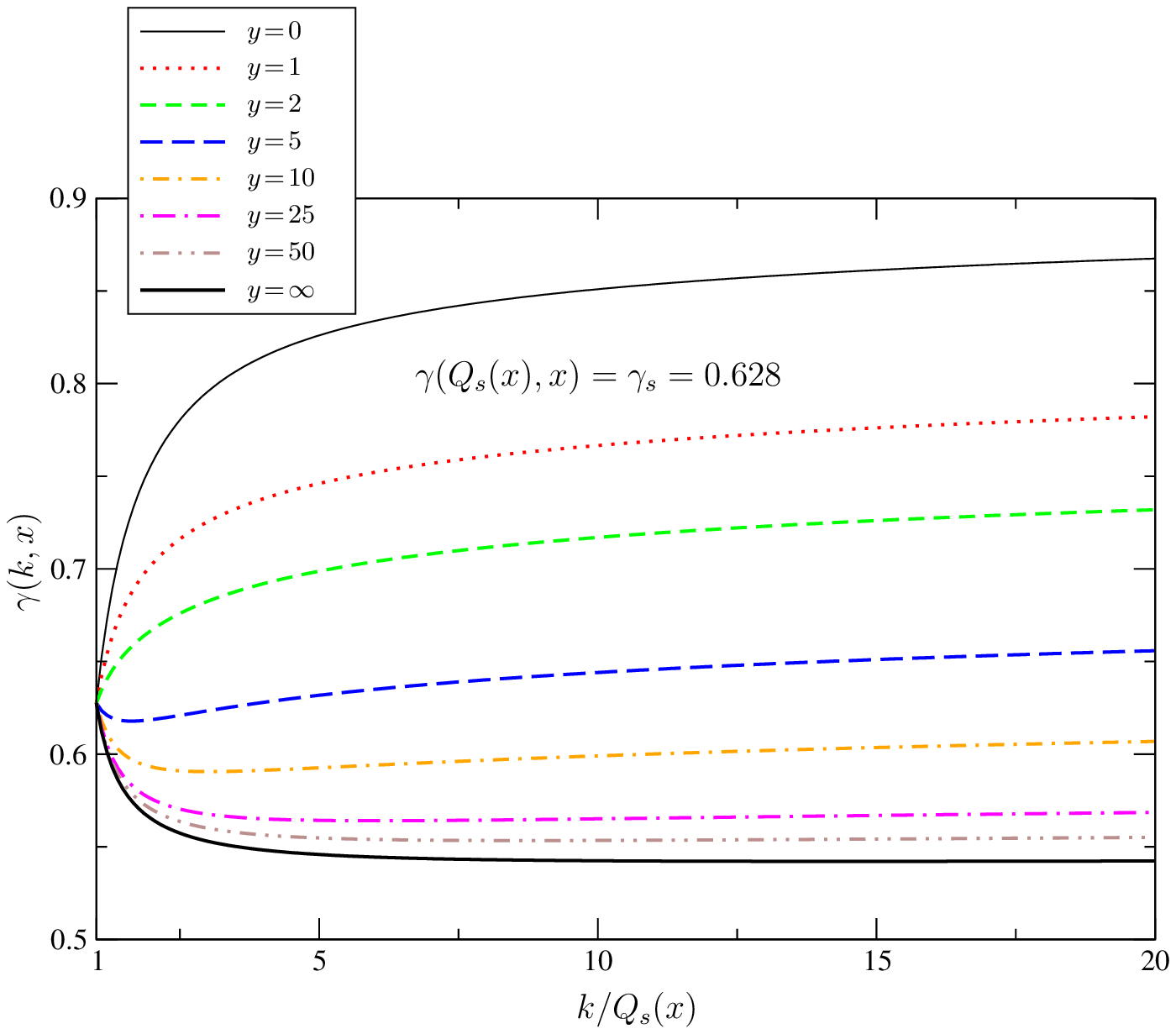}
\caption{\small\label{gamma_k_fig} $\gamma(k,x)$ as a function of
  $k/Q_s(x)$ for various rapidities $y=\log x_0/x$.}
\end{figure}

As mentioned before, DHJ actually considered Eq.\ (\ref{Ngamma})
with $\gamma(r,x)$ replaced by $\gamma(1/k,x)$
\cite{KKT,DHJ1,DHJ2}. This approximation scheme we will discuss next. 
The procedure of extracting $\gamma$ becomes quite different when $\gamma$
depends on $k$, since the dipole cross section $N$ then
depends on both $r$ and $k$, so that it is not related to
$\mathcal{N}(k,x)$ by a straightforward inverse Fourier transform
(\ref{Fourier}) anymore:  
\begin{equation}
 \mathcal{N}(k,x) \equiv \int_0^\infty \frac{dr}{r}\:{
   J}_0(kr)\,\lf(1-\exp\lf[-\frac{1}{4}(r^2Q^2_s(x))^{\gamma(k,x)}\rg]\rg)\,.
\label{Nk2}
\end{equation}
Instead of by the inverse Fourier transform, we will extract
$\gamma$ by trying to numerically solve Eq.\ (\ref{Nk2}), imposing
the following condition. In order to test the Ansatz of DHJ, we
will fix $\gamma(k,x)$ in such a way that it equals the constant
$\gamma_s$ at the saturation scale:
\begin{equation}
 \gamma(Q_s(x),x)=\gamma_s\approx 0.628\,.
\label{gs1}
\end{equation}
This determines the dependence of $Q_s$ on $x$, which can be found by
explicitly solving
\begin{equation}
\label{Qsdef2}
 \mathcal{N}(Q_s(x),x)= 
\int_0^\infty \frac{{d}z}{z}\:{
   J}_0(z)\,\lf(1-\exp\lf[-\frac{1}{4}(z^2)^{\gamma_s}\rg]\rg)\approx
0.19 \,.
\end{equation}
Now we can extract $\gamma$ from relation (\ref{Nk2}) for any
given value of $x$ and $k$. Fig~\ref{gamma_k_fig} shows the
results for $\gamma(k,x)$ as a function of $k/Q_s$ in the EGS
region, for a broad range of rapidities. For small rapidities the
resulting $\gamma$ looks very similar to the one of DHJ
(cf.\ Fig.\ 4 of Ref.\ \cite{DHJ2}). This gives us an indication
of how the DHJ distribution would behave under BK evolution. We reiterate
that one cannot use the DHJ distribution itself as a starting distribution as
it is not defined over the whole $r$ or $k$ range. As one
can see, for larger $y$ the resulting $\gamma$ is not compatible 
with the DHJ parameterization (\ref{gammaparam}) anymore; it first decreases 
($\Delta \gamma <0$) before it rises towards 1 asymptotically.  

\begin{figure}[hbt]
\centering
\includegraphics*[width=110mm]{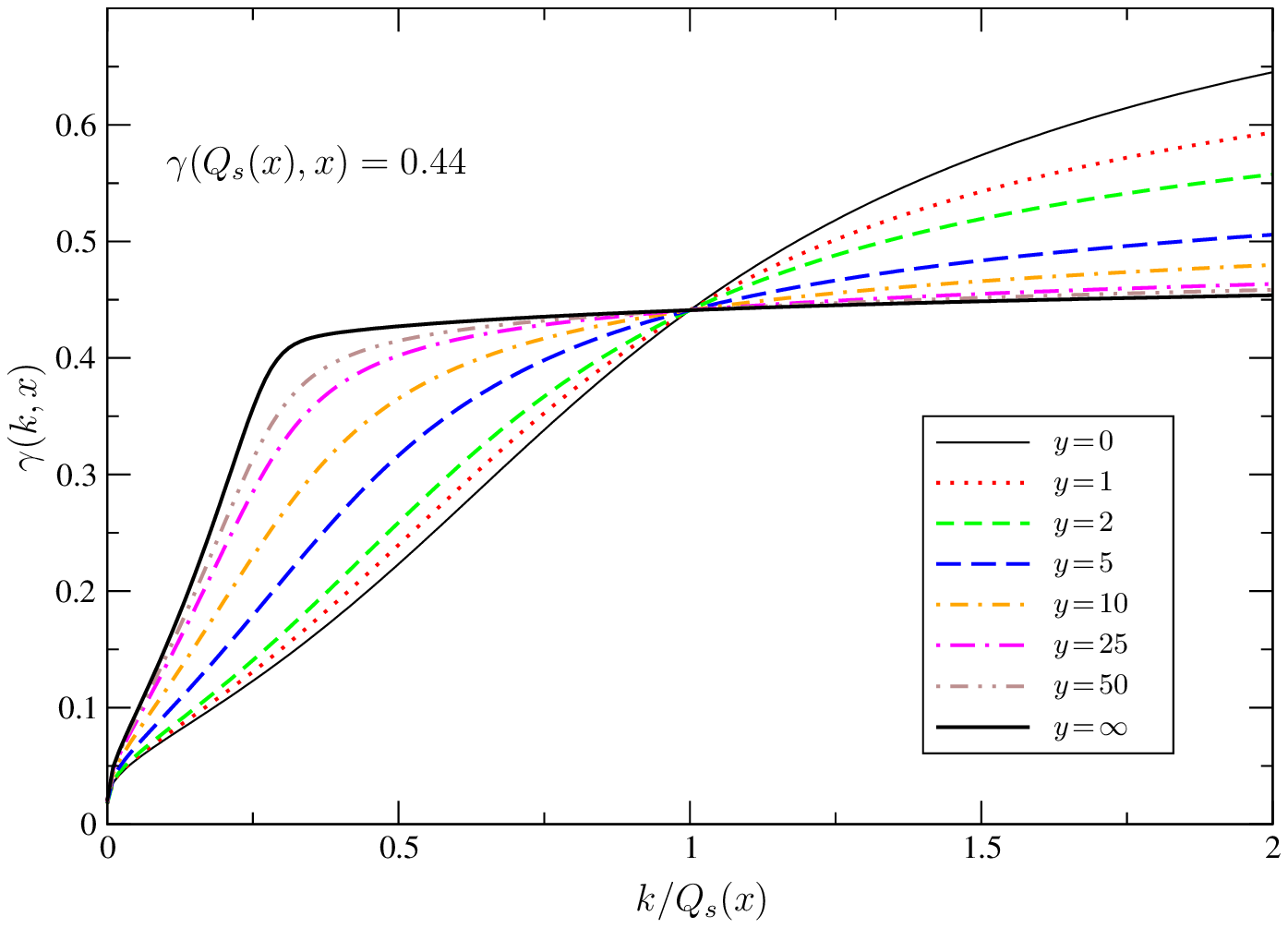}
\caption{\small\label{gamma_k_gammalow_fig} $\gamma(k,x)$ as a
 function of $k/Q_s(x)$ for various rapidities $y=\log x_0/x$.}
\end{figure}

In Fig.\ \ref{gamma_k_gammalow_fig} the same result is shown but now
for a smaller choice of constant $\gamma$ at the saturation scale,
$\gamma(k=Q_s,x)=0.44$. This value is the small-$x$ limit
(\ref{gammars}) of
$\gamma(r=1/Q_s(x),x)$ in the approach of the previous subsection. 
In this case $\Delta
\gamma$ remains positive above the saturation scale 
and the result looks very much like the DHJ
parameterization for all $y$. A fit of $\gamma$ as given in Eq.\
(\ref{gammaparam}) to the numerical results for $k/Q_s=1,2,\ldots,5$ and
rapidities $y=1,2,\ldots,5$ (in order to allow for a comparison to
DHJ's results) yields $\lambda \approx 1$ and $d \approx 3$, although
it must be emphasized that the shapes of the curves are not exactly of
the form (\ref{gammaparam}). 

\subsection{The saturation region}

The DHJ Ansatz was specifically intended to describe the dipole
scattering amplitude in the EGS region, but it is nevertheless
interesting to see what $\gamma$ will look like if one were to assume
the same form of $N$ in the saturation region.  One may expect the
solution to be geometrically scaling in the region $r \geq 1/Q_s$,
which would be the case if $\gamma(r,x)=\gamma(r Q_s(x))$. A constant
$\gamma$ would of course be a special case. However, the numerical
solution of the BK equation does not display geometric scaling, not
even in the saturation region, although for large rapidities ${\cal
N}(k,x)$, as a function of $k/Q_s(x)$, converges to a well-defined
limit.  The same conclusion applies to $\gamma$ upon assuming the
Ansatz (\ref{Ngamma}) for $N(r,x)$. This is clearly seen in Fig.\
\ref{gammar1_fig}.

The situation is similar if we replace $\gamma(r,x)$ with
$\gamma(1/k,x)$, keeping $\gamma$ at the saturation scale $k=Q_s$
fixed, with the exception that in some cases no solutions for $\gamma$
can be found at all in the saturation region. The existence of a
solution in a certain range of $k/Q_s$ depends on the chosen
saturation scale or, in the above discussed approach, equivalently, on
the fixed value of $\gamma$ at the saturation scale, which determines
the saturation scale. In fact, we find that if $\gamma(k=Q_s(x),x)$ is
fixed to equal $\gamma_s=0.628$, the Ansatz (\ref{Ngamma}) is not
compatible with the solution of the BK equation in the range $k/Q_s
\approx 0.2 - 0.4$, where no value of $\gamma(k,x)$ satisfies the
relation (\ref{Nk2}) for large rapidities. If instead we fix $\gamma$
at the saturation scale to equal the limit (\ref{gammars}), the
saturation scale is shifted towards smaller values, so that there is a
unique solution for $\gamma$ for all $k$ and $x$. Above the saturation
scale $\gamma(k,x)$ is in agreement with the type of parameterization
of DHJ (\ref{gammaparam}). For asymptotic rapidities, $\gamma$
approaches a limiting curve, which is geometrically scaling for all
$k$ values, not just at the saturation scale (which was the case by
construction). Note that below the saturation scale the solution is
far from constant, although for large rapidities the solution is
rather constant for a considerable momentum range below $Q_s$. But
eventually, for very small $k$, it drops significantly to a small
nonvanishing value. We emphasize that this is the case under the
assumption that the form (\ref{Ngamma}) also holds for momenta below
the saturation scale.  Let us mention that in the region where
$\gamma$ eventually drops towards smaller values $r\,Q_s(x)$ becomes
effectively already so large that $N(r,x)\approx1$, so that the exact
value of $\gamma$ eventually becomes irrelevant, see
Fig.~\ref{ncal_fig}b.

\subsection{Saturation scale dependence on $x$}

\begin{figure}[htb]
\centering
\includegraphics*[width=110mm]{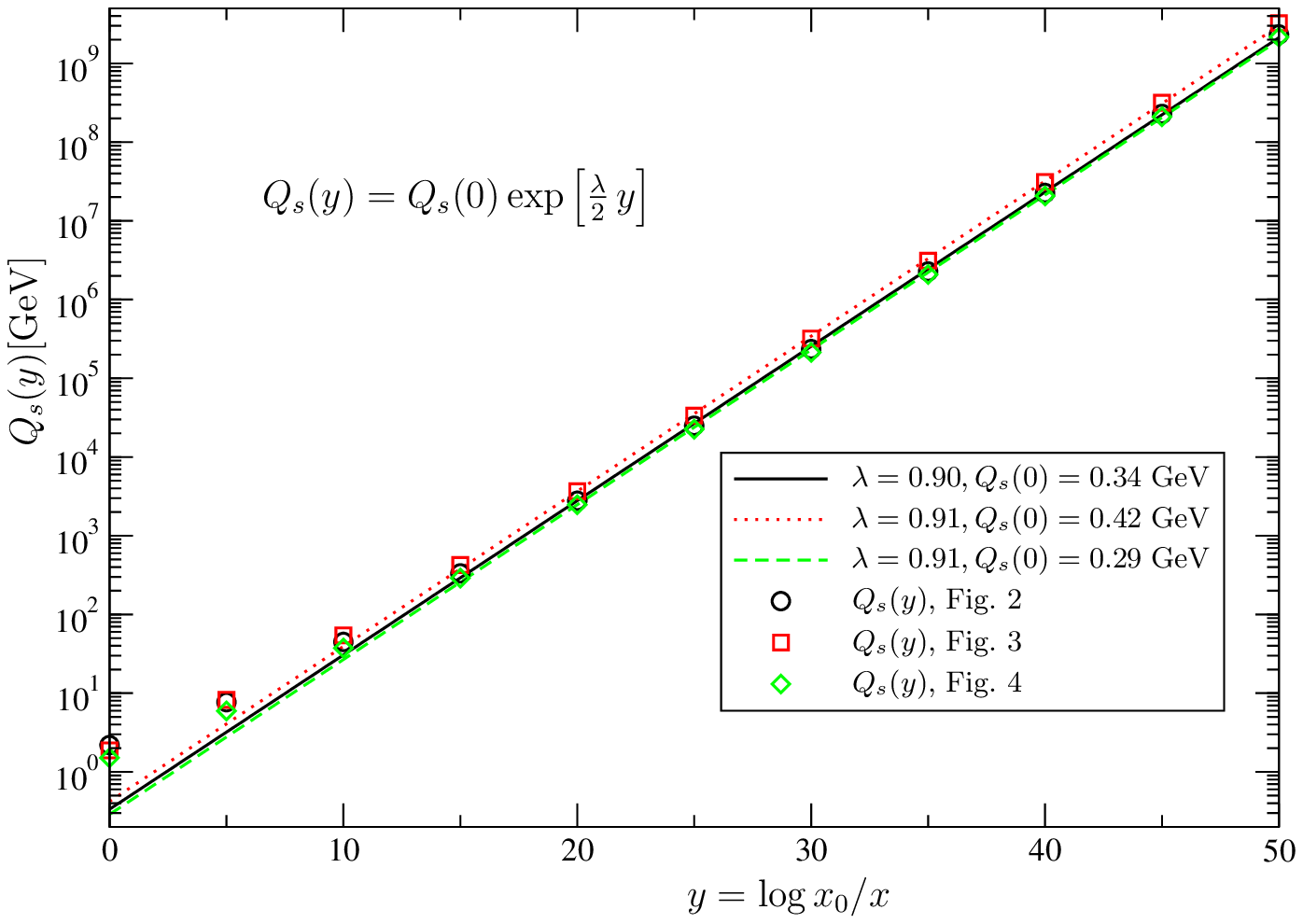}

\caption{\small\label{Qs_fig} The saturation scale from the various
  approaches for $y=\log x_0/x=0,5,10,\ldots ,50$ (symbols) and the
  related fits of the form $Q_s(y)=Q_s(0)\exp[\lambda\,y/2]$ in the
  fixed coupling case, where $\bar{\alpha}_s=0.2$.}
\end{figure}

In Fig.\ \ref{Qs_fig} we display the values of $Q_s(x)$ for a fixed
coupling $\bar{\alpha}_s=0.2$ as a function of $y=\log x_0/x$, in
order to show that (at least for larger rapidities, $y>10$ say) the
behavior $Q_s(x) \sim x^{-\lambda/2}$ is found in all scenarios
considered here. The values of the parameters can be compared with the
GBW values of $\lambda \simeq 0.3$ and $Q_s(y=0) \simeq 1$ GeV. We
note that strictly speaking one has to include a factor 2/3 in
$Q_s(y=0)$ if the DHJ Ansatz for $N_F$ is used for calculating
$Q_s(x)$. This is the case in e.g.\ Eqs.\ (\ref{Qsdef1}) and
(\ref{Qsdef2}). The larger value of $\lambda \approx 0.9$ is in
agreement with the asymptotic result $\lambda\approx
\bar{\alpha}_s\chi(\gamma_s)/\gamma_s\approx \bar{\alpha}_s\,4.88$
from LO BFKL equation with saturation boundary conditions
\cite{MuellerTr} and $\lambda=4\bar{\alpha}_s$ from the DLA approach
\cite{LevinTuchin,Levin:1999mw,Golec-Biernat:2001if,IIM2}.
 
We have also investigated the case where the coupling constant runs
with the energy scale, which is given by the inverse of the relevant
dipole size. For details, we refer to \cite{Enberg:2005cb,BKSolver}. 
The energy scale turns out to be effectively of the order of the
resulting saturation scale $Q_s$. The running of the coupling results
in a change in the dependence of $Q_s(x)$ on $x$, for which we now
find approximately $Q_s(y)=Q_s(0)\exp[c\,\sqrt{y}/2]$, which in fact
is equally well supported by all relevant DIS data
\cite{Gelis:2006bs}.  Since we have plotted every result for the
anomalous dimensions as a function of dimensionless quantities, i.e.\
$rQ_s(x)$ or $k/Q_s(x)$, none of those results is significantly
affected by including running in the coupling constant. It has to be
mentioned that since the BK equation is of leading order in
$\alpha_s$, including the running coupling is of course not a
consistent treatment of higher order effects. For that reason we
mainly focus on the fixed coupling case.

\subsection{Dependence on the initial conditions}
\label{initial_sec}

To illustrate the dependence of some of our results on the choice of a
starting distribution, we replace the MV distribution with the
following choice:
\begin{equation}
N(r,x=x_0)=1-\exp\lf[-\frac{1}{4}(r^2Q_s^2(x=x_0))^{\gamma_0}\rg]\,.
\label{Ansatz3}
\end{equation}

\begin{figure}[htb]
\centering
\includegraphics*[width=86mm]{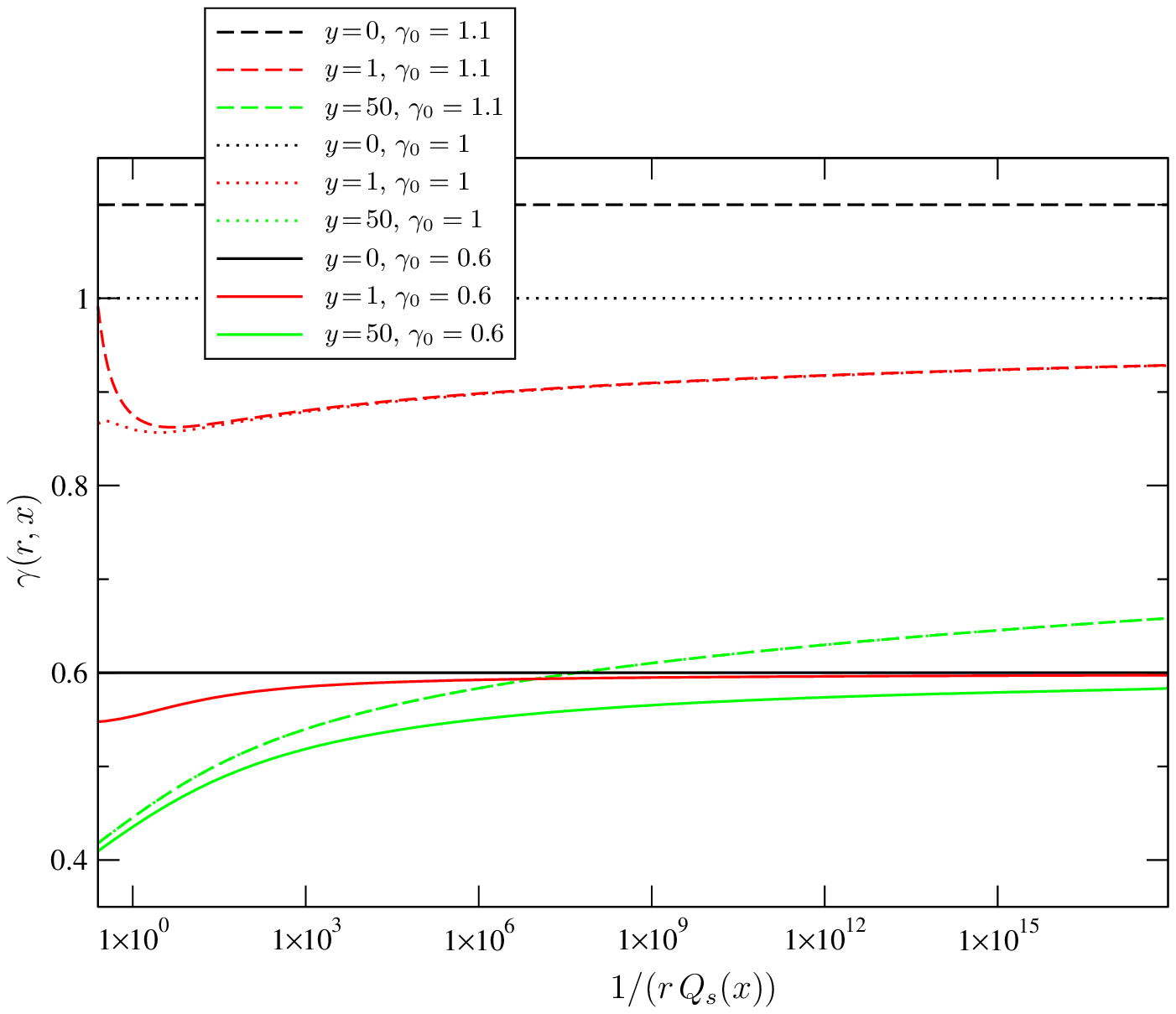}
\includegraphics*[width=86mm]{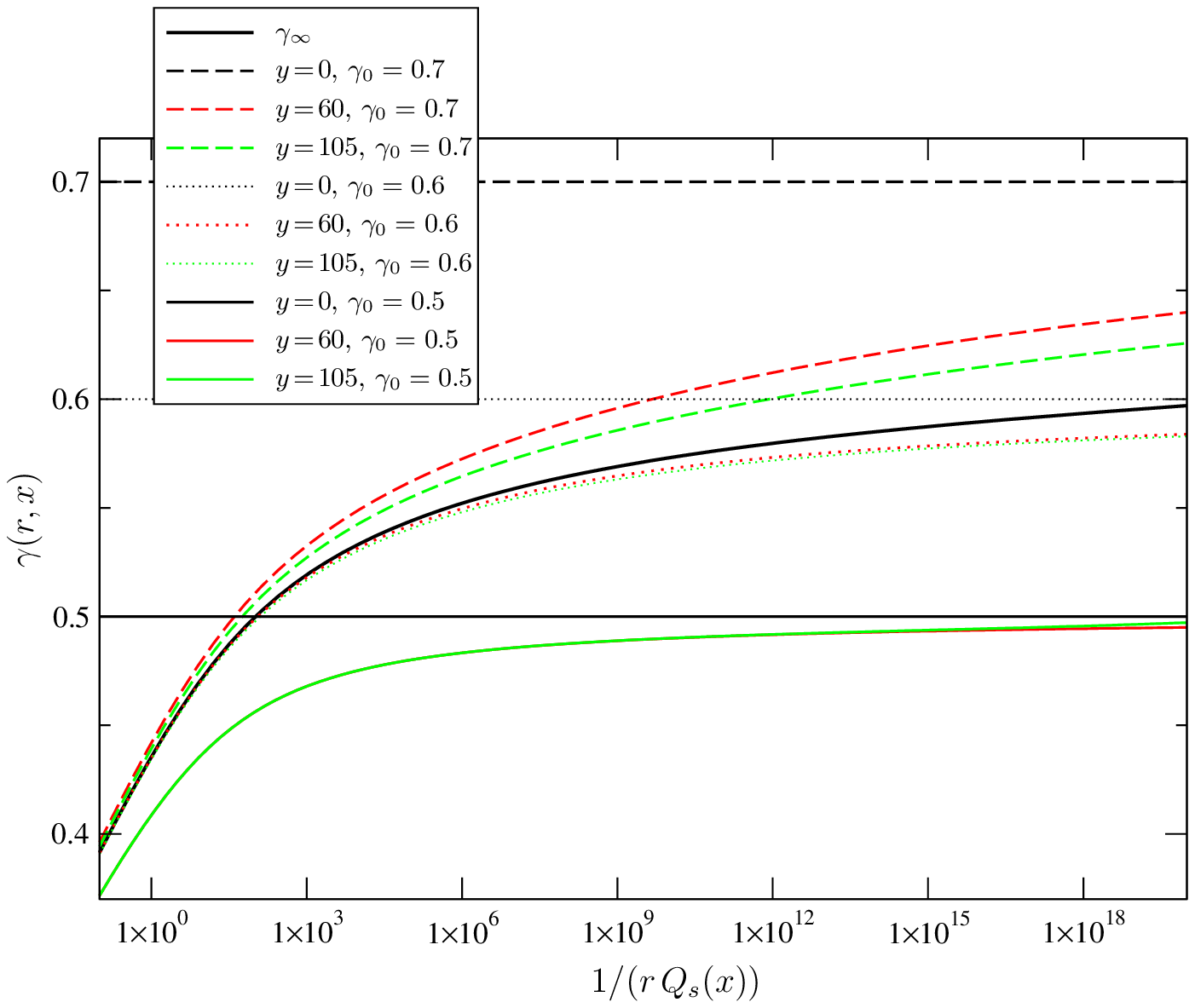}
\caption{\small\label{ini_fig} a) $\gamma(r,x)$ for three different
  rapidities and initial conditions $\gamma_0=0.6,1,1.1$. To
  illustrate the small $r$ asymptotics, we show $\gamma(r,x)$ over a
  very large range. Note that the curves with $\gamma_0=1$ and $1.1$
  for $y=50$ are hard to distinguish.
  \\
  b) $\gamma(r,x)$ as a function of $1/(r Q_s(x))$ and initial
    conditions $\gamma_0=0.5, 0.6$ and 0.7 for large rapidities
    $y=\log x_0/x$. Here, the curves for $y=60$ and $y=105$ largely coincide for
    both $\gamma_0=0.5$ and 0.6.} 
\end{figure}

Fig.~\ref{ini_fig}a shows the resulting $\gamma(r,x)$ as a function of
$1/(r Q_s(x))$ for the initial conditions $\gamma_0=0.6, 1$ and $1.1$,
in a very large $r$ range in order to display the asymptotic small-$r$
behavior. As expected from the discussion in Sec.~\ref{sect_BK}, in
the limit of very large $1/(r Q_s)$ at finite rapidity, $\gamma$
approaches either the initial condition $\gamma_0$, if $\gamma_0 < 1$,
or the saddle point, which is 1 in this limit.

Fig.~\ref{ini_fig}b shows a similar plot for the initial conditions
$\gamma_0=0.5, 0.6$ and 0.7 at larger rapidities.  For $\gamma_0=0.5$
in the large-$y$ limit $\gamma$ approaches a limiting function that
monotonically rises towards $\gamma_0$ at small $rQ_s$. The limiting
function for $\gamma_0=0.6$ behaves similarly.  For $\gamma_0=0.7$ the
limiting function coincides to good approximation with the one of
$\gamma_0=0.6$ up to small values of $rQ_s$, but becomes slightly
larger as $rQ_s$ decreases further. In the entire investigated range
the limiting function of $\gamma_0=0.7$ coincides with the one of the
MV starting distribution (\ref{Ansatz2}), partly shown in
Fig.~\ref{gammar1_fig}. We conclude that if $\gamma_0(rQ_s)$ is
sufficiently large, one always reaches a universal limiting function
$\gamma_\infty(rQ_s)$. This function equals approximately 0.44 at the
saturation scale and approaches a limit $\gamma_\infty(rQ_s \to 0)$
which seems to be slightly larger than 0.6. As discussed below, the
latter limit seems consistent with the theoretical expectation of
0.628. In the specific case of constant $\gamma_0$ the universal
limiting curve is always reached if $\gamma_0$ is larger than
$\gamma_\infty(rQ_s \to 0)$.

A similar dependence on $\gamma_0$ has been observed in Ref.\
\cite{Munier:2003sj} on the traveling wave solution. This solution
depends on a parameter $\gamma$ which equals either $\gamma_s$ or
$\gamma_0$, whichever is the smallest. In our approach the role of
$\gamma_s$ of the traveling wave is played by the function
$\gamma_\infty$. At large $k/Q_s$ and $y$ the traveling wave solution
is dominated by the term $(k^2/Q_s^2(x))^{-\gamma}$, leading to the
previously mentioned expectation that $\gamma_\infty(rQ_s \to 0) =
\gamma_s$.

Finally, we compare our results to those of Ref.\
\cite{Albacete:2004gw}, especially Fig.~4. There it is found that for
small $rQ_s$ the scattering amplitude can be described by $N(r)=a (r
Q_s)^{2\gamma}(\log(r Q_s)+\delta)$, where $\gamma$ approaches
approximately $0.65$ for large $y$, regardless of the initial
condition.  We find that the particular value of 0.65 is consistent
with our result for $\gamma_\infty$, if we take into account the
different functional form for $N(r,x)$ and the specific range of $r$
and $y$ in which $\gamma$ was calculated.  However, even for
$\gamma_0=c/2=0.42$, Ref.\ \cite{Albacete:2004gw} find no dependence
of $\gamma$ on the initial condition, which seems to disagree with our
results and those of Ref.\ \cite{Munier:2003sj}.

\section{Conclusions}

The numerical solutions of the BK equation do not display exact
geometric scaling, although they approach a solution showing such
scaling at asymptotic $y$. Assuming the solutions to be of the form
(\ref{Ngamma}), where scaling violations are encoded in the
``anomalous dimension'' $\gamma$, therefore leads to the conclusion
that $\gamma(r,x)$ is not a function of $rQ_s(x)$ exclusively. In
particular, it is never simply a constant, not even at the saturation
scale ($r=1/Q_s$). At asymptotically large rapidities, $\gamma$
reaches a limiting function $\gamma_\infty(rQ_s(x))$. This function is
universal for a large range of initial conditions.  At the saturation
scale, this function equals approximately $0.44$, which is
considerably smaller than the corresponding values of $\gamma$ in the
phenomenological models \cite{IIM,KKT,DHJ2}. For small values of
$rQ_s$ the limiting function seems to reach $\gamma_s$, in accordance
with the traveling wave results of Refs.\
\cite{Munier:2003vc,Munier:2003sj}. This is not in contradiction with
the behavior of $\gamma$ in the limit of $rQ_s(x)\to 0$ at any fixed
$x$, where $\gamma$ is determined by the starting distribution or the
BFKL saddle point. It merely implies that the two limits $rQ_s(x)\to
0$ and $x\to 0$ do not commute.

Although the numerical solutions of the BK equation lead to a
$\gamma(r,x)$ that is never a constant as a function of $x$ at the
saturation scale, performing the replacement of $\gamma(r,x) \to
\gamma(1/k,x)$ does allow one to find a solution for which
$\gamma(k=Q_s,x)$ is kept fixed. The behavior of $\gamma(1/k,x)$ is
then qualitatively similar to the DHJ parameterization for small
rapidities. However, the usually considered choice
$\gamma(k=Q_s,x)=\gamma_s=0.628$ yields some unwanted features, i.e.\
the fact that $\Delta \gamma <0$ in a region above the saturation
scale and the absence of solutions below the saturation scale,
although the Ansatz was not intended for that region. Keeping
$\gamma(k=Q_s,x)$ fixed at a smaller value, e.g.\ at
$\gamma_\infty(rQ_s=1) \approx 0.44$, seems more suitable, but it
remains to be investigated whether such a choice allows for a good fit
of all relevant DIS, $d$-$Au$ and $p$-$p$ data. One might expect this
to be possible, as the KKT model \cite{KKT}, which has $\gamma=0.5$ at
the saturation scale, is able to describe the $d$-$Au$ data.

The resulting saturation scales in the various approaches we adopted,
evolve quite similarly with decreasing $x$, namely approximately as
$Q_s(x) = Q_s(x_0) (x_0/x)^{\lambda/2}$, albeit with somewhat
different normalizations $Q_s(x_0)$. In the case of a running coupling
constant, the dependence on $y$ changes from $\exp (\lambda y/2)$ to
$\exp (c \sqrt{y}/2)$, which is equally well supported by all relevant
DIS data \cite{Gelis:2006bs}.

Our conclusions about the DHJ model apply as well to the IIM and KKT
models in the EGS region. It would be interesting to consider
modifications of these models for the dipole scattering amplitude that
are compatible with both the BK equation and the data. Given the fact
that the BK evolution does not respect geometric scaling around $Q_s$,
phenomenological parameterizations that reflect this feature would
seem a natural choice. Fortunately, the LHC and a possible future
electron-ion collider will provide data over a larger range of momenta
and rapidities, so that one can expect to test the evolution
properties of the models more accurately.
 
\begin{acknowledgments}
  D.B.\ thanks Adrian Dumitru, Fran\c{c}ois Gelis, Arata Hayashigaki,
  Dima Kharzeev, Kazunori Itakura, Larry McLerran, Kirill Tuchin and
  Raju Venugopalan, for sharing their knowledge on this topic with him
  on several occasions.  We thank Adrian Dumitru for helpful comments
  on the manuscript. We thank one of the referees for helping to
  clarify the role of the initial condition and for pointing out the apparent
  discrepancy of our results with those of Ref.\
  \cite{Albacete:2004gw}.  This research is part of the research
  program of the ``Stichting voor Fundamenteel Onderzoek der Materie
  (FOM)'', which is financially supported by the ``Nederlandse
  Organisatie voor Wetenschappelijk Onderzoek (NWO)''.
\end{acknowledgments}

\end{document}